\documentclass[12pt]{article}
\usepackage{amssymb}
\usepackage{epsfig}

\catcode`\@=11
\def\numberbysection{\@addtoreset{equation}{section}
        \def\theequation{\thesection.\arabic{equation}}}

\def\beq{\begin{equation}}
\def\eeq{\end{equation}}
\numberbysection
\begin{document}
\begin{titlepage}
\begin{center}
\hfill  \\
\vskip 1.in {\Large \bf Snyder Geometry and Quantum Field Theory} \vskip 0.5in P. Valtancoli
\\[.2in]
{\em Dipartimento di Fisica, Polo Scientifico Universit\'a di Firenze \\
and INFN, Sezione di Firenze (Italy)\\
Via G. Sansone 1, 50019 Sesto Fiorentino, Italy}
\end{center}
\vskip .5in
\begin{abstract}
We find that, in presence of the Snyder geometry, the notion of translational invariance needs to be modified, allowing a momentum dependence of this symmetry. This step is necessary to build the maximally localized states and the Feynman rules of the corresponding quantum field theory.
\end{abstract}
\medskip
\end{titlepage}
\pagenumbering{arabic}
\section{Introduction}

At the Planck scale we expect that the gravitational effects of the test particle's energy modify the structure of space-time, inducing a finite limit $\Delta x_0$ on the possible resolution of distances. String Theory, the leader candidate to describe the quantum gravitational phenomena, suggests also a certain type of correction to the uncertainty relation \cite{1}-\cite{5}

\beq \Delta x \Delta p \geq \frac{\hbar}{2} ( 1 + \beta ( \Delta p )^2 ) \ \ \ \ \beta > 0 \label{11} \eeq

implying a non-zero minimal uncertainty $\Delta x_0$. This uncertainty relation can be modeled by introducing a correction term to the commutation relations:

\beq [ \hat{x}, \hat{p} ] = i \hbar ( 1 + \beta \hat{p}^2 ) \label{12} \eeq

In the case of $n$-dimensions, the simplest generalization of this algebra gives rise to the Snyder geometry \cite{6}. Snyder introduced it as a lattice regularization of space-time, but his interpretation is misleading. In fact in the momentum space there is no limitation of any kind, while the fuzziness $\Delta x_0$ is reduced to the space-time coordinates.

In this article we study in detail the implications that this algebra induces on quantum field theory. Firstly we need to modify the translational invariance, which is broken by this algebra, in order to make covariant with it.

This concept allows us to build the maximally localized states which carry the maximal quantum information compatible with this algebra. These states are then necessary to compute the basic Feynman rules, as the vertex and the propagator. In particular the propagator $G(\xi-\eta)$ is no more singular in the limit $\xi\rightarrow\eta$, avoiding the $UV$ divergencies, typical of quantum field theory.

\section{The algebra}

Snyder algebra is simply obtained modifying the representation of the position operator in the following way ( for simplicity we work with an Euclidean metric  )

\beq \hat{x}^\mu \ = \ i \hbar \ [ \ \delta^{\mu\nu} + \beta p^\mu p^\nu \ ] \ \frac{\partial}{\partial p^\nu} \label{21} \eeq

from which we obtain the following commutation relations:
\begin{eqnarray}
& \ & [ \hat{x}^\mu, \hat{p}^\nu ] \ = \ i \hbar \ [ \ \delta^{\mu\nu} + \beta \ \hat{p}^\mu \hat{p}^\nu \ ] \nonumber \\
& \ & [ \hat{x}^\mu, \hat{x}^\nu ] \ = \
 i \hbar \beta \ [ \ \hat{p}^\nu \hat{x}^\mu  - \hat{p}^\mu \hat{x}^\nu \ ] \label{22}
\end{eqnarray}

These define a possible $n$-dimensional generalization of the unique modification of $1d$
quantum mechanics

\beq [ \hat{x}, \hat{p} ] \ = \ i \hbar [ 1 + \beta \hat{p}^2 ] \label{23}\eeq

inducing a non zero minimal uncertainty on the position measure

\beq {( \Delta x )}^2 \ \geq \ \beta \hbar^2 \label{24} \eeq

In the Euclidean $n$-dimensional case, each spatial dimension contributes with an analogous uncertainty giving rise to the following formula

\beq {( \Delta x )}^2 \ \geq \ n \beta \hbar^2 \label{25} \eeq

Therefore the physical meaning of the Snyder geometry is introducing a spatial hypercube of side-length $\sqrt{\beta} \hbar$ which is not accessible to the position measurements.
It results in a very efficient regularization of the $UV$ divergencies of quantum field theory.

\section{ The $\rho$ variables}

It is convenient introducing a new representation of the Snyder algebra, given in terms of a new variable $\rho^\mu$ living on a compact hypersphere of radius $\frac{1}{\sqrt{\beta}}$ :

\begin{eqnarray}
& \ & \hat{x}^\mu \ = \ i \hbar \ \sqrt{ 1 - \beta \rho^2 } \ \frac{\partial}{\partial \rho^\mu} \ \ \ \ \ \ \ \ \ 0 < \rho^2 < \frac{1}{\beta} \nonumber \\
& \ & \hat{p}^\mu \ = \ \frac{\rho^\mu}{\sqrt{ 1 - \beta \rho^2 }} \label{31}
\end{eqnarray}

This representation will help us in the following to simplify all the calculations. It works also in the $1d$ case:

\begin{eqnarray}
& \ & \hat{x} \ = \ i \hbar \ \sqrt{ 1 - \beta \rho^2 } \frac{\partial}{\partial \rho} \ \ \ \ \ \ \ \ \ 0 < \rho < \frac{1}{\sqrt{\beta}} \nonumber \\
& \ & \hat{p} \ = \ \frac{\rho}{\sqrt{ 1 - \beta \rho^2 }} \label{32}
\end{eqnarray}

where it faithfully reproduces all the results reported in \cite{7}.

The operators $\hat{x}^\mu$ and $\hat{p}^\mu$ are symmetric with respect to the following scalar product

\beq < \psi(\rho) |  \phi(\rho) > \ = \ \int d^n \rho \ \frac{\psi^{*}(\rho) \ \phi(\rho)}{\sqrt{ 1 - \beta \rho^2 }} \label{33} \eeq

In the momentum representation, the operator $\hat{x}^\mu$ is again symmetric with the following choice:

\beq \hat{x}^\mu \ = \ i \hbar \ \left[ \frac{\partial}{\partial p^\mu} \ + \ \beta p^\mu p^\nu \frac{\partial}{\partial p^\nu} \ + \ \beta \left( \frac{n+1}{2} \right) p^\mu
\right] \label{34} \eeq

The corresponding scalar product is now

\beq < \psi(p) | \phi(p) > \ = \ \int \ d^n p \ \psi^{*}(p) \phi(p) \label{35} \eeq

The operator $\hat{x}^\mu$ doesn't admit a unique self-adjoint extension, which is of course a consequence of the minimal uncertainty of the position measurement.

\section{Modifying the translation invariance}

Snyder geometry is compatible with the rotation group but not with the translation one.
The fact that the coordinates $\hat{x}^\mu$ do not commute is an obstacle to the study of maximally localized states compatible with it. This is simply because the eigenvalue equation

\beq \hat{x}^\mu \ \psi_\xi ( \rho ) \ = \ \xi^\mu \ \psi_\xi ( \rho ) \label{41} \eeq

doesn't make sense, since the coordinates $\hat{x}^\mu$ cannot be simultaneously diagonalized, and the combination $ ( \hat{x}^\mu - \xi^\mu ) $
is not covariant with respect to the commutation rules.

Our contribution is modifying the concept of translation in order to make it covariant with respect to the Snyder algebra and to make sense of the eigenvalue equation for $\hat{x}^\mu$.

Our strategy is to define a momentum dependent operator analogous to the translation

\beq \hat{x}^\mu_\xi \ = \ \xi^\mu \ f(\rho^2) \ + \ \rho^\mu \ ( \rho \cdot \xi ) \ g ( \rho^2) \label{42} \eeq

imposing that the combination

\beq \hat{\tilde{x}}^\mu \ = \ \hat{x}^\mu - \hat{x}^\mu_\xi \label{43} \eeq

satisfies to the same commutation rules of the Snyder algebra. Moreover we require that for $n=1$ this operator reduces to the simple translation:

\begin{eqnarray}
& \ & 2 f'(\rho^2) \ - \ g (\rho^2) \ + \ \frac{\beta}{1-\beta\rho^2} \ f( \rho^2 ) \ = \ 0 \nonumber \\
& \ & f(\rho^2) \ + \ \rho^2 \ g(\rho^2) \ = \ 1 \label{44} \end{eqnarray}

This system can be resolved by introducing the series

\begin{eqnarray}
& \ & f(\rho^2) \ = \ 1 + \sum_{n=1}^{\infty} \ \beta^n f_n (\rho^2)^n \nonumber \\
& \ & \rho^2 \ g(\rho^2) \ = \ \sum_{n=1}^{\infty} \ \beta^n g_n (\rho^2)^n \label{45}
\end{eqnarray}

from which

\beq f_n \ = \ - \ g_n \ = \ - \ \frac{(2n-2)!!}{(2n+1)!!} \label{46} \eeq

These series can be resummed to give:

\beq f(\rho^2) \ = \ 1 \ - \ \rho^2 \ g(\rho^2) \ = \ 1 \ + \frac{1}{2} \ \int^1_0 \ dx \ \ln [ 1 - 4 x ( 1-x) \beta \rho^2 ] \label{47} \eeq

An alternative representation for $f(\rho^2)$ is

\beq f(\rho^2) \ = \ \sqrt{\frac{1-\beta\rho^2}{\beta\rho^2}} \ \arcsin \sqrt{\beta\rho^2} \label{48} \eeq

Coming back to the momentum variables we obtain

\beq f(p^2) \ = \ \frac{1}{\sqrt{\beta p^2}} \ \arctan \sqrt{\beta p^2} \ = \
\int^1_0 \ dx \ \frac{1}{ 1 + \beta p^2 x^2 } \label{49} \eeq

from which we can give the final representation for the translation operator

\beq \hat{x}^\mu_\xi \ = \ \int^1_0 \ dx \ \left( \frac{\xi^\mu \ + \ \beta p^\mu ( p \cdot \xi ) x^2 }{ 1 \ + \ \beta p^2 x^2 } \right) \label{410} \eeq

For $n=1$ it is obvious that $ \hat{x}^\mu_\xi \rightarrow \xi$.

By construction the combination $\hat{x}^\mu - \hat{x}^\mu_\xi$ is covariant with respect to the commutation rules. Now we want to show that the following analogue of the eigenvalue equation

\beq \hat{x}^\mu \ \psi_\xi \ = \ \hat{x}^\mu_\xi \ \psi_\xi \ \ \ \ \ \ \ \ \ < \psi_\xi | \psi_\xi > \ = \ 1 \label{411} \eeq

can be solved. In fact the solution in the momentum coordinates is simply

\beq \psi_\xi (p) \ = \ \frac{c}{( 1+\beta p^2 )^{\frac{n+1}{4}}} \ e^{ \frac{\xi \cdot p}{i \hbar \sqrt{\beta p^2}}  \ \arctan \sqrt{\beta p^2} } \label{412} \eeq

that, translated in the $\rho$ variables, results in

\beq \psi_\xi (\rho) \ = \ \left[ \left( \frac{\beta}{\pi} \right)^{\frac{n}{2}} \frac{\Gamma(\frac{n+1}{2})}{\sqrt{\pi}} \right]^{\frac{1}{2}} \
e^{ \frac{\xi \cdot \rho}{i \hbar \sqrt{\beta \rho^2}}  \ \arcsin \sqrt{\beta \rho^2} } \label{413} \eeq

Let us note that the solution (\ref{412}) is the obvious generalization for $n\neq 1$ of the state studied in \cite{7}. We will need this wave function to define the maximally localized states and the analogue of the Fourier transform.

Obviously the scalar product of two wave functions with two independent translations is different from zero as in the case $n=1$ , due to the fuzziness of space-time:

\beq < \psi_\xi | \psi_\eta > \ = \ c^2 \ \int \ \frac{d^n \rho}{\sqrt{1-\beta \rho^2}}
\ e^{ \frac{(\xi -\eta)\cdot \rho}{i \hbar \sqrt{\beta \rho^2}}  \ \arcsin \sqrt{\beta \rho^2} } \label{414} \eeq

The angular part can be computed

\beq \int \ d \Omega \ e^{ - i \frac{\alpha \cdot \rho}{\sqrt{\rho^2}}  \ \arcsin \sqrt{\beta \rho^2} } \ = \ (2\pi)^{\frac{n}{2}} \ \frac{ J_{\frac{n}{2}-1} ( \alpha \arcsin \sqrt{\beta \rho^2} ) }{
( \alpha \arcsin \sqrt{\beta \rho^2} )^{\frac{n}{2} -1}} \label{415} \eeq

where $\alpha = \frac{|\xi-\eta|}{\hbar \sqrt{\beta}}$ and $J_\nu$ is the Bessel function of the first kind.

Introducing the following change of variables $\rho = \frac{1}{\sqrt{\beta}} \ \sin x$
the radial part of the integral (\ref{414}) results in :

\beq < \psi_\xi | \psi_\eta > \ = \ \left( \frac{2}{\alpha} \right)^{\frac{n}{2}} \frac{\alpha}{\sqrt{\pi}} \ \Gamma \left( \frac{n+1}{2} \right) \ \int^{\frac{\pi}{2}}_0 \ dx \ \frac{ ( sin x )^{n-1} }{ x^{\frac{n}{2}-1} } \ J_{\frac{n}{2}-1} ( \alpha x ) \label{416} \eeq

Unfortunately this integral cannot be computed for generic $n$. For $n=1$ we re-obtain

\beq < \psi_\xi | \psi_\eta > \ = \ \frac{sin \left[ \frac{\pi}{2} \alpha \right]
}{ \frac{\pi}{2} \alpha } \label{417} \eeq

and for $n=3$

\beq < \psi_\xi | \psi_\eta > \ = \ \frac{1}{\pi \alpha} \left[ - Si \left[ \frac{\pi}{2} (\alpha-2) \right] + 2 Si \left[ \frac{\pi}{2} \alpha \right] -
Si \left[ \frac{\pi}{2} (\alpha+2) \right] \right] \label{418} \eeq

\section{ Maximally localized states }

After modifying the concept of translation, we are ready to build the maximally local states, i.e. the physical states with which we will build the Feynman rules of the theory. To do this it is necessary studying the following equation:

\beq ( \ \hat{x}^\mu - \ \hat{x}^\mu_\xi  \ + \ i \ k \ \hat{p}^\mu \ ) \ \psi_k ( p^2 ) \ = 0 \label{51} \eeq

choosing the parameter $k$ in order to minimize the uncertainty associated to
$\psi_k (p^2)$. For simplicity we resolve this problem in the origin and then we apply the translation wave function discussed in the previous section to define the generic case.

In the origin the equation (\ref{51}) written in the variables $\rho$ implies:

\beq \frac{\partial}{\partial \rho^\mu} \ \psi_k (\rho^2) \ = \ - \frac{k}{\hbar} \ \frac{\rho^\mu}{(1-\beta \rho^2)} \ \psi_k(\rho^2) \ \ \ \ \ \ \ < \psi_k | \psi_k > = 1
\label{52} \eeq

that is resolved by

\beq \psi_k (\rho^2) \ = \ \left[ \ \left( \frac{\beta}{\pi} \right)^{\frac{n}{2}} \ \frac{ \Gamma \left( \frac{k}{\beta\hbar} \ + \ \frac{n+1}{2} \right) }{
\Gamma \left( \frac{k}{\beta\hbar} \ + \ \frac{1}{2} \right) } \right]^{\frac{1}{2}} \
( 1-\beta \rho^2 )^{\frac{k}{2\beta \hbar}} \label{53} \eeq

The indetermination on this state results in

\beq ( \Delta x )^2|_k \ = \ < \psi_k | \hat{x}^2 | \psi_k > \ = \ \frac{n}{2\beta} \ \frac{k^2}{\left( \frac{k}{\beta\hbar} \ - \ \frac{1}{2} \right)} \ \ \ \ \rightarrow \ k \ = \ \beta \hbar \label{54} \eeq

The minimum is obtained for $ k = \beta \hbar $ and the value coincides with what discussed at the beginning of the article:

\beq ( \Delta x )^2|_{\beta\hbar} \ = \ n \beta \hbar^2 \label{55} \eeq

The translated wave function can be obtained simply by multiplying it with the solution to the modified eigenvalue problem (\ref{411})

\beq \psi^{ml} (\rho) \ = \ \left[ \ \left( \frac{\beta}{\pi} \right)^{\frac{n}{2}} \ \frac{ \Gamma \left( \frac{n+3}{2} \right) }{
\Gamma \left( \frac{3}{2} \right) } \right]^{\frac{1}{2}} \
( 1-\beta \rho^2 )^{\frac{1}{2}} \ e^{ \frac{\xi \cdot \rho}{i \hbar \sqrt{\beta \rho^2}}  \ \arcsin \sqrt{\beta \rho^2} } \label{56} \eeq

This is our final wave function describing the maximal localization in the neighborhood of the point $\xi^\mu$.

\section{ Feynman rules}

The maximally localized states are the building blocks to define the Feynman rules of the quantum field theory based on the Snyder algebra. In particular we need to compute the vertex and the propagator ( see also \cite{8} )

\begin{eqnarray}
& \ & \tilde{\delta} ( \xi^{ml}, \eta^{ml} ) \ = \ < \xi^{ml} | \eta^{ml} > \nonumber \\
& \ & G( \xi^{ml}, \eta^{ml} ) \ = \ \frac{\hbar^2}{(\Delta x_0)^2} \ < \xi^{ml} | \frac{1}{ p^2 + m^2 } | \eta^{ml} > \ \ \ \ \ (\Delta x_0)^2 = n \beta \hbar^2 \label{61} \end{eqnarray}

The vertex is just the scalar product of two maximally localized states:

\beq \tilde{\delta} ( \xi^{ml}, \eta^{ml} ) \ = \ c^2 \int^{\frac{1}{\sqrt{\beta}}}_0 \
d \rho \rho^{n-1} \ \sqrt{1-\beta\rho^2} \ \int \ d \Omega \ e^{ - i \frac{\alpha \cdot \rho}{\sqrt{\rho^2}}  \ \arcsin \sqrt{\beta \rho^2} } \label{62} \eeq

where

\beq c^2 \ = \ \left( \frac{\beta}{\pi} \right)^{\frac{n}{2}} \ \frac{\Gamma\left(\frac{n+3}{2}\right)}{\Gamma\left(\frac{3}{2}\right)} \ \ \ \
| \alpha | \ = \ \frac{|\xi-\eta|}{\hbar\sqrt{\beta}} \label{63} \eeq

The final result is the integral

\beq \tilde{\delta} (\xi^{ml}, \eta^{ml} )  \ = \ \left( \frac{2}{\alpha} \right)^{\frac{n}{2}} \ \frac{ \Gamma \left( \frac{n+3}{2} \right) }{ \Gamma \left( \frac{3}{2} \right) } \ \alpha
\ \int^{\frac{\pi}{2}}_0 \ dx \ \left[ \frac{ ( sin x )^{n-1} - ( sin x )^{n+1} }{ x^{\frac{n}{2}-1} } \right] \ J_{\frac{n}{2}-1} ( \alpha x ) \label{64} \eeq

For $n=1$ it is exactly solvable in terms of elementary functions

\beq \tilde{\delta} (\xi^{ml}, \eta^{ml} )|_{n=1}  \ = \ \frac{ \sin \frac{\pi}{2} \alpha }{ \pi }  \left[ \frac{1}{ (\frac{\alpha}{2}) - (\frac{\alpha}{2})^3 }
\right] \label{65} \eeq

confirming what is computed in \cite{7}. The propagator is instead:

\beq G( \xi^{ml}, \eta^{ml} ) \ = \ \frac{1}{n \beta} \ < \xi^{ml} | \frac{1}{p^2+m^2} | \eta^{ml} > \label{66} \eeq

Due to the following identity

\beq \frac{1}{p^2+m^2} \ = \ \frac{1}{(1-\beta m^2)^2} \ \frac{1}{\rho^2 + \frac{m^2}{1-\beta m^2}} \ - \ \frac{\beta}{1-\beta m^2} \label{67} \eeq

the propagator can be divided in two contributions

\beq G( \xi^{ml}, \eta^{ml} ) \ = \ G_0 ( \xi^{ml}, \eta^{ml} ) \ - \ \frac{1}{n ( 1- \beta m^2)} \ \tilde{\delta} (\xi^{ml}, \eta^{ml}) \label{68} \eeq

The core term is obviously

\beq G_0 (\xi^{ml},\eta^{ml}) \ = \ \left( \frac{2}{\alpha} \right)^{\frac{n}{2}} \ \frac{ \Gamma \left( \frac{n+3}{2} \right) }{ \Gamma \left( \frac{3}{2} \right) } \ \frac{\alpha}{n ( 1-\beta m^2 )^2 }
\ \int^{\frac{\pi}{2}}_0 \ dx \ \left[ \frac{ ( sin x )^{n-1} - ( sin x )^{n+1} }{ x^{\frac{n}{2}-1} \left( sin^2 x + \frac{\beta m^2}{1-\beta m^2} \right)  } \right] \ J_{\frac{n}{2}-1} ( \alpha x ) \label{69} \eeq

This integral cannot be computed for $m^2 \neq 0$. In the massless case we obtain instead

\beq G_0 (\xi^{ml},\eta^{ml})|_{m^2=0} \ = \ \left( \frac{2}{\alpha} \right)^{\frac{n}{2}} \ \frac{ \Gamma \left( \frac{n+3}{2} \right) }{ \Gamma \left( \frac{3}{2} \right) } \ \frac{\alpha}{n }
\ \int^{\frac{\pi}{2}}_0 \ dx \ \left[ \frac{ ( sin x )^{n-3} - ( sin x )^{n-1} }{ x^{\frac{n}{2}-1} } \right] \ J_{\frac{n}{2}-1} ( \alpha x ) \label{610} \eeq

In the limit $\alpha \rightarrow 0$ ( i.e. $\xi\rightarrow \eta$ and $\beta$ fixed )
we obtain the interesting result

\beq G_0 ( \xi^{ml}, \eta^{ml} )|_{m^2=0} \ ( \alpha \rightarrow 0 ) \ = \ \frac{n+1}{n(n-2)} \label{611} \eeq

and for the complete propagator

\beq G ( \xi^{ml}, \eta^{ml} )|_{m^2=0} \ ( \alpha \rightarrow 0 ) \ = \ \frac{3}{n(n-2)} \label{612} \eeq

From this formula we note that the propagator, as defined in (\ref{61}), makes sense only for
$n \geq 3$, while for $n=2$ the integral diverges logarithmically. In the limiting case
$n=3$ for $\alpha \rightarrow 0$ we obtain simply the identity, while for $\alpha \rightarrow \infty$ ( $\beta \rightarrow 0$, $| \xi - \eta |$ = fixed ) the propagator assumes the usual form

\beq G ( \xi^{ml}, \eta^{ml} )|_{m^2=0} \ ( \alpha \rightarrow \infty ) \ \propto \ \frac{1}{|\xi-\eta|} \label{613} \eeq

The physical meaning of this result is interesting. The $UV$ divergencies of quantum field theory arise because the propagator is a singular function in the limit
$\xi \rightarrow \eta$, i.e. it is a distribution due to the locality property.

Usually in the perturbative expansion we find products of propagators and the products of distributions are generally ill defined, as in the following case

\beq [G ( \xi-\eta )]^2 \ = \ \int \ \frac{d^4p}{(2\pi)^4} \ e^{i p \cdot (\xi-\eta)} \ \int \ \frac{d^4k}{(2\pi)^4} \  \frac{1}{(k^2 - m^2)((p-k)^2 - m^2)} \label{614} \eeq

and we need to introduce subtractions to make sense of the perturbative result. Instead in the case of the Snyder geometry the propagator is finite in the limit
$\xi\rightarrow\eta$, and the product of propagators is a well defined function.
The theory is regularized in its roots ( modifying the quantization rules ) without the need of introducing extra subtractions.

\section{Conclusions}

In summary, the Snyder algebra represents the simplest generalization of $1d$ quantum mechanics with a non-zero minimal position uncertainty. It introduces a hypercube of side-length $\sqrt{\beta}\hbar$, not accessible to the measurements.

We have succeeded to modify the translational invariance in order to make it covariant with the non-commutative algebra. This step is fundamental to define the maximally localized states that realize the highest physical information accessible.

Thanks to these states we can build the Feynman rules like the vertex and the propagator.
Unlike ordinary quantum field theory, the propagator is no more a distribution, i.e. singular in the limit $\xi\rightarrow\eta$, but finite. This property assures that the $UV$ divergencies of quantum field theory are absent and there is no need of ad hoc subtractions.

Then we discuss the limits of our technique. The first limit we see is that our scheme is strongly dependent on our Euclidean choice of the space-time metric. Frankly we don't know if our formulas can be analytically continued to the Minkowskian case \cite{9}. The second limit is that the introduction of a non-locality of space-time is in conflict with the principle of unitarity ( which is strictly related to the locality principle ). Since with the fuzziness of space-time we have no control of the Planck scale at the measurement level, we expect that the requirement of unitarity must be limited to the laboratory energy scale. Work is in progress in this direction.

\end{document}